\documentclass[10pt,letterpaper]{article}
\usepackage[top=0.85in,left=2.75in,footskip=0.75in]{geometry}
\usepackage{comment}
\usepackage{verbatim}
\usepackage{amsmath,amssymb}

\usepackage{changepage}

\usepackage[utf8x]{inputenc}

\usepackage{textcomp,marvosym}

\usepackage{cite}

\usepackage{nameref,hyperref}

\usepackage[right]{lineno}

\usepackage{microtype}
\DisableLigatures[f]{encoding = *, family = * }

\usepackage[table]{xcolor}

\usepackage{array}

\newcolumntype{+}{!{\vrule width 2pt}}

\newlength\savedwidth

\usepackage{caption}
\usepackage{subcaption}

\raggedright
\usepackage[aboveskip=1pt,labelfont=bf,labelsep=period,justification=raggedright,singlelinecheck=off]{caption}

\makeatletter
\renewcommand{\@biblabel}[1]{\quad#1.}
\makeatother

\usepackage{lastpage,fancyhdr,graphicx}
\usepackage{epstopdf}
\fancyheadoffset[L]{2.25in}
\fancyfootoffset[L]{2.25in}
\usepackage{multirow}

\newcommand{\bin}{\beta_{N,S\rightarrow S,S} }
\newcommand{\gin}{\beta_{S,Q \rightarrow Q,Q} }
\newcommand{\din}{\beta_{S,N \rightarrow Q,N}}
\newcommand{\oin}{\beta_{Q,S \rightarrow S,S}}

\newcommand{\bspon}{\delta_{N \rightarrow S}}
\newcommand{\gspon}{\delta_{S \rightarrow Q}}
\newcommand{\ospon}{\delta_{Q \rightarrow S}}

\usepackage[colorinlistoftodos]{todonotes}\reversemarginpar

\makeatletter\newcommand\subsubsubsection{\@startsection{paragraph}{4}{\z@}{-2.5ex\@plus -1ex \@minus -.25ex}{1.25ex \@plus .25ex}{\normalfont\normalsize\bfseries}}
\newcommand\subsubsubsubsection{\@startsection{subparagraph}{5}{\z@}{-2.5ex\@plus -1ex \@minus -.25ex}{1.25ex \@plus .25ex}{\normalfont\normalsize\bfseries}}
\makeatother

\begin{document}

\vspace*{0.2in}

\begin{flushleft}
{\Large
\textbf\newline{Improving tobacco social contagion models using agent-based simulations on networks} 
}
\newline
\\
Adarsh Prabhakaran\textsuperscript{1*},
Valerio Restocchi\textsuperscript{1},
Benjamin D. Goddard\textsuperscript{2}
\\
\bigskip
\textbf{1} Artificial Intelligence and its Applications Institute, School of Informatics, University of Edinburgh, Edinburgh, United Kingdom 
\\
\textbf{2} School of Mathematics and Maxwell Institute for Mathematical Sciences, University of Edinburgh, Edinburgh, United Kingdom
\bigskip

* a.prabhakaran@sms.ed.ac.uk

\end{flushleft}
\section*{Abstract}

Over the years, population-level tobacco control policies have considerably reduced smoking prevalence worldwide. However, the rate of decline of smoking prevalence is slowing down. Therefore, there is a need for models that capture the full complexity of the smoking epidemic. These models can then be used as test-beds to develop innovative new policies to decrease the spread of smoking. Current models of smoking dynamics mainly use ordinary differential equation (ODE) models, where studying the effect of an individual's contact network is challenging. They also do not consider all the interactions between individuals that can lead to changes in smoking behaviour, implying that the current models do not consider valuable information on the spread of smoking behaviour.

In this context, we develop an agent-based model (ABM), calibrate and then validate it on historical trends observed in the US and UK regions. Our ABM considers spontaneous terms, interactions between agents, and the agent's contact network. To explore the effect of the underlying network topology on smoking dynamics, we test the ABM on six different networks, both synthetic and real-world. In addition, we also compare the ABM on each network with an ODE model. Our results suggest that the dynamics from the ODE model are similar to the ABM only when the network structure is fully connected (FC). The FC network performs poorly in replicating the empirical trends in the data, while the real-world network best replicates it amongst the six networks. Further, when information on the real-world network is unavailable, our ABM on Lancichinetti-Fortunato-Radicchi benchmark networks (or networks with a similar average degree as the real-world network) can be used to model smoking behaviour. These results suggest that networks are essential for modelling smoking behaviour and that our ABM can be used to develop network-based intervention strategies and policies for tobacco control.

\section*{Introduction}

Smoking is one of the leading preventable causes of death, disability and disease in the US \cite{surgeonsreport2014,surgeonsreport2020} and in the UK \cite{cornish2019adult} and it is one of the most significant avoidable hazard factors for cancer \cite{banks2015tobacco} and respiratory diseases \cite{ferkol2014global}. Not only is smoking a global health burden, but also an economic burden, which significantly outweighs the economic benefits from its production and sales  \cite{drope2018tobacco}.

Acknowledging the need for active efforts on tobacco control in 2005, 182 countries ratified the first international public health treaty, the WHO Framework Convention on Tobacco Control (FCTC) \cite{world2004framework}. This involved cigarette taxation, smoke-free zones, public media campaigns \cite{vallone2009socioeconomic}, restrictions on advertisement for cigarettes \cite{gilpin1997trends,rose2013peer}, health warnings, cessation support \cite{brown2014equity} and control on access to tobacco products \cite{millett2011increasing,schneider2011happens}. Effective implementation of such tobacco control policies has decreased deaths, extended lifespan (maximum number of years a person can live) \cite{holford2014tobacco}, increased life expectancy (estimate of years a person can expect to live) \cite{van2012estimating} and is also associated with a predicted decrease in healthcare expenditure \cite{lightwood2016smoking}. The success of the tobacco control policies can be clearly seen in the decline in smoking prevalence over the past 35 years, where the global prevalence decreased from over 40.0\% in the 1980s to 22.8\% in 2020 worldwide \cite{ng2014smoking,world2019global,world2021global}.

However, despite this progress in tobacco control, the rate of decline of smoking prevalence has slowed down, and the world's economies still spend more than one trillion USD per year on smoking-related health expenditures and loss of productivity \cite{acharya2016economics,goodchild2018global}. 

Most of the strategies and policies developed focus on population-level interventions such as marketing and pricing \cite{caryl2021simulating}. However, the decline in the rate of decrease of smoking prevalence suggests that the effectiveness of these strategies might have reached a saturation point. This must be tackled with new strategies and effective policies that address social contagion. To develop such strategies, we need a new model to act as a test-bed.

Such a model, should be able to fully capture the complexity of the spreading nature of smoking behaviour. Therefore, it is necessary to understand the phenomenon in detail beyond just the standard population-level.

A large body of evidence suggests that tobacco use can spread through social ties and, depending on the social tie, the chance of both smoking initiation and quitting changes \cite{christakis2008collective,blok2017role}. 
Other studies have mainly focused on the influence of peers on school-going adolescents \cite{ennett2008peer} and the effects of homophily on smoking initiation \cite{mercken2009dynamics}. In addition, studies have shown that smoking cessation is more likely when smokers are part of a non-smoker group  \cite{go2010peer}.
Similarly, relapsing into smoking is more probable if the individual is part of a smoker group \cite{blok2017role}. Detailed longitudinal network analyses have also shown patterns like groups of smokers quitting together and a higher probability of smoking cessation when a neighbour in an individual's social network quits \cite{christakis2008collective}. 

However, current models of smoking do not fully incorporate social contagion dynamics. Traditionally, smoking is compared to a disease spreading in a population due to its contagion-like behaviour \cite{sharomi2008curtailing, zaman2011optimal,zaman2011qualitative,zaman2017dynamics}. This similarity allows the use of tools from epidemiology to model the propagation of smoking behaviour. However, epidemiology has moved on to flexible agent-based models (ABM), but tobacco control has not. Therefore, models in tobacco control cannot account for the network effects that have been observed from empirical analysis due to the following reasons.

First, these models do not account for the topology of the network formed through social ties. When modelling any social contagion, it is known that the social interaction between individuals plays an important role \cite{hodas2014simple,shin2022social}. Therefore such models can only be accurate if they account for the underlying social network of the population. 
Even though compartmental models try to incorporate social interactions between groups of individuals, there is no network structure involved. To reduce the complexity and improve explainability these models ignore the network topology of the system and instead assume a homogeneous well-mixed population which means that any individual can infect others in the system \cite{anderson1992infectious,kermack1927contribution,kermack1932contributions,kermack1933contributions}. This means that vital information from the actual social network is not taken into consideration \cite{moore2000epidemics}. 

Second, when these models do consider networks, they arbitrarily fix the network structure. The models on smoking behaviour which consider networks are agent-based models on networks \cite{chao2015dynamic,schaefer2013social,schaefer2012dynamic}. In these cases, the networks are usually arbitrarily fixed as a scale-free network or on small scaled school-network. Since real-world social contact networks of adults can be very different from synthetic networks and school networks, there is a need for careful characterisation of the smoking behaviour of these models on different network topologies.

Finally, even though empirical research shows that smoking depends on interactions with social ties \cite{christakis2008collective}, not all of these interactions have been considered for modelling the spread of smoking.

 Although one of the first theoretical studies which modelled smoking cessation advocates the use of interactions between smokers and quitters \cite{castillo1997mathematical}, interactions which lead to smoking cessation and relapse are still not used in the models \cite{schaefer2013social}. Consequently, in these models smoking cessation and relapse are usually determined only by spontaneous terms \cite{sharomi2008curtailing, zaman2011optimal,zaman2011qualitative,zaman2017dynamics}.

A closely related field to the spread of smoking behaviour is opinion dynamics, wherein mathematical and computational models are used to study the spread of opinions in a population by considering social influence. However, unlike opinions, smoking is a health hazard, which makes the spread of the smoking behaviour very similar to that of an infectious disease.Therefore, modelling smoking behaviour can be seen through a hybrid lens of epidemiology and opinion dynamics.  

Over the years, the effect of social networks on individual and population behaviour have been studied in both opinion dynamics and epidemiology. Multiple methods and approaches have been designed to study the spread of diseases and opinion according to the situation and amount of information available.  More recently, these models have been used to study a variety of social contagion including  obesity \cite{hill2010infectious}, emotions \cite{hill2010emotions}, alcoholism \cite{lee2010optimal,sharma2015analysis}, substance abuse \cite{white2007heroin}, information spreading \cite{zhou2020realistic} to name a few. Due to the similarity of spread of smoking behaviour to other social contagion, we can use insights from these fields to develop models.

We use the ABM approach to model the spread of smoking behaviour. ABM is a class of computational techniques which relies on dynamical interactions between autonomous agents to understand the macroscopic consequences of the system due to such local interactions. We use ABM for the following reasons:

First, unlike the ODE models, ABMs can easily be extended to multiple theoretical and real-world networks. This versatile nature allows it to be applied to different population structures and more realistic models of the system. Therefore the effect of network structure on the smoking dynamics can easily be studied by changing the underlying network topology in ABMs. Consequently, we characterise the dynamics of smoking behaviour in multiple different network topologies. 

Next, interactions between agents can easily be incorporated into ABMs without significantly increasing the complexity of the model. Conversely, when multiple interactions are considered in an ODE model, the model becomes complex and difficult to solve analytically. By constructing ODE models which are not solvable (or ones in which the solutions are complex and lengthy), it becomes difficult to validate and analyse the nature of the solutions analytically. Instead, such ODE systems must be solved using numerical methods to approximate the solutions.

In our ABM for smoking, we include three state change processes in the ABM: smoking initiation, cessation and relapse. Each of these processes can occur due to both interactions and spontaneously. Thus accounting for multiple possible interactions which can lead to a change in smoking behaviour.

Finally, ABMs are flexible enough to become effective test-beds for developing new policies. One of the main applications of studying contagion (both infectious diseases and social contagion) is to develop strategies to contain them. Effective strategies and interventions may stop never-smokers from picking up smoking, motivate smokers to quit and stop former smokers from relapsing. Due to the contagious nature of smoking, we can potentially use network-based strategies developed from studies on infectious diseases and other social contagions. 

 First, we develop an agent-based model, which considers multiple possible interactions along with spontaneous terms to study the spread of smoking. Our model can be used to develop network-based intervention strategies and policies for tobacco control. Furthermore, we show the need for ABM on networks by comparing the dynamics with a traditional ODE model.  
Second, we show that the ABM for smoking behaviour on an Fully Connected (FC) network shows similar dynamics to that of an ODE model. Additionally, we show that ABM on FC and ODE models should not be used to model smoking behaviour as they replicate the real-world data with poor accuracy compared to the other networks.
Next, we explore the effect of the underlying network topology on smoking dynamics. We find that the underlying network structure affects smoking dynamics considerably. However, synthetic networks with the same average-degree, reproduced the historic data and showed similar characteristics as that of the real-world networks. Specifically, we show that Lancichinetti Fortunato Radicchi benchmark networks and random network networks can be used to develop intervention strategies when complete information on underlying network topology of a local population is not available.

\section*{Methods}

This section describes the model structure, data, and the networks used, along with the modelling choices involved in developing the ABM for smoking behaviour.


We highlighted the need for a tobacco control model to incorporate both spontaneous and interaction terms, as well as to consider appropriate network topologies. To achieve this, we use a synthetic population of $n$ (please note the use of small case $n$, since $N$ has already been used to denote Never-smokers) agents in an undirected and unweighted network $G$. To show the effects of the network topology, we make the agents interact on six different networks (described later in the section) and compare the observed dynamics.

\subsection*{Description of the agent-based model}        
In our model, each agent can be in one of the following smoking states: Never-Smoker ($N$), Smoker ($S$) or Quitter ($Q$). An agent is a never-smoker if they have never smoked before, while an agent who smokes any tobacco product daily or occasionally falls into the smoker state. Finally, if a smoker quits smoking even temporarily, they are labelled a quitter. 


To make models more realistic over long time periods, epidemiological models usually include vital dynamics. Traditionally, this is done by including constant mortality and birth rates in the equations. In network-based models, a constant birth rate and mortality rate can lead to older agents having a higher number of social contacts, thus increasing their influence on other agents. However, in reality, the number of social connections does not increase with age but instead peaks in the mid-twenties and then decreases with age \cite{bhattacharya2016sex}. However, tackling these problems would significantly increase the complexity of the model since the model will have to include age-dependent mortality and birth rates, and the network generation process will have to be tweaked to ensure that the network retains its properties while adding and removing agents. Hence, we do not include vital dynamics in the model and run our experiments in time frames where the vital dynamics can be ignored. We calibrate and validate our model in time periods of less than 30 years to minimise the effect of vital dynamics.


\begin{figure}
    \centering
    \includegraphics[scale=0.4]{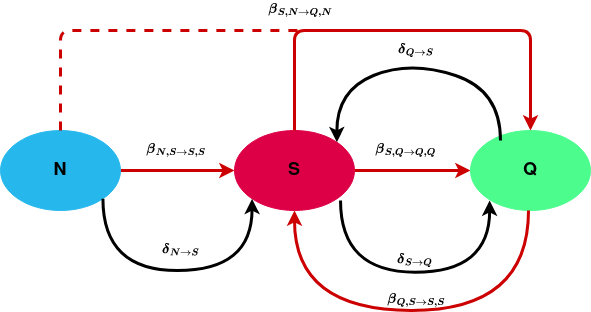}
    \caption{The figure shows the schematic representation of the state change processes involved in the ABM. The interaction parameters are represented by the red arrows, while the black arrows show the spontaneous terms in the schematic. All three state-change processes are shown in the figure. First, an N-agent can initiate smoking spontaneously ($\bspon$) or due to the interaction with an S-agent ($\bin$). Similarly, an S-agent can quit spontaneously ($\gspon$) or due to interaction with other non-smoker agents (Q-agent: $\gin$ or N-agent: $\din$ ). Like the other processes, Q-agents relapse into smoking spontaneously ($\ospon$) or due to interaction ($\oin$) with an S-agent.}
    \label{fig:my_label}
\end{figure}

\subsubsection*{State change processes}   
        We incorporate three main processes into the model: smoking initiation, smoking cessation and relapse into smoking. The three state change processes involve interaction-based state changes as well as spontaneous state transitions. We assume that each exposure to an agent with a different smoking status is independent of the previous exposure. We then use a binomial approximation to compound the effect of multiple independent interactions on one agent simultaneously to calculate the probability of state change. In Supplementary section \nameref{Sup_binomial}, we provide a detailed description and derivation of the expressions used.   

        

        

\paragraph*{Smoking Initiation:}
First and foremost, we define the transition of a never-smoker into a smoker as smoking initiation. In the model, an N-agent can initiate smoking in two ways. First, through a random probability $\bspon$ depicting various external influences like advertisements, movies, and the presence of tobacco shops influencing an N-agent to pick up smoking. Second, through interactions with other S-agents in its network neighbourhood with a probability $\bin$. 
We use the binomial approximation mentioned before to calculate the expression in \ref{eq_smoke_init}, which gives the probability of smoking initiation due to interaction. 

        \begin{equation}
           P  =  \frac{ n_S}{n} (1- (1- \bin)^{n_S})
           \label{eq_smoke_init}
       \end{equation} 
        

\paragraph*{Smoking Cessation:}
Following smoking initiation, we define the process of an S-agent quitting smoking as smoking cessation. Similar to smoking initiation, smoking cessation can also happen in two ways. First, due to various external influences like mass-media campaigns, cigarette box advertisements, and higher taxes. This external influence is incorporated into the model through the spontaneous term $\gspon$. Second, due to interactions with non-smokers. However, both Q-agents as well as N-agents fall under the non-smoker category. Therefore unlike the other state change processes, interactions with both the other states can lead to smoking cessation. $\din$ represents the probability of cessation of an S-agent due to N-agents in its network neighbourhood. At the same time, the probability of cessation due to other Q-agents in its network neighbourhood is given by $\gin$. Subsequently, the probability of smoking cessation due to interaction of an S-agent with multiple Q-agents is given by \ref{eq_quit_Q} while \ref{eq_quit_N} gives the same probability but due to interaction with multiple N-agents. 

        \begin{equation}
           P  =  \frac{ n_Q}{n} (1- (1- \gin)^{n_Q})
           \label{eq_quit_Q}
       \end{equation} 
        
        \begin{equation}
           P  =  \frac{ n_N}{n} (1- (1- \din)^{n_N})
           \label{eq_quit_N}
       \end{equation}

\paragraph*{Smoking Relapse:}
Finally, we define picking up smoking after a period of abstinence as a smoking relapse. 
Smoking relapse is similar to smoking initiation, except that the Q-agent gets influenced instead of an N-agent. Similar to the other two cases, smoking relapse can happen in two ways. $\ospon$ represents the probability of a Q-agent relapse into smoking due to external influence. Additionally, $\oin$ represents the probability of Q-agent relapsing into smoking due to interaction with its immediate network-neighbour S-agents.
Here, like the smoking initiation, only interactions with other S-agents can cause a Q-agent to relapse into smoking. The probability of Q-agent relapsing due to its interaction with multiple S-agents is shown in \ref{ew_relapse}.
        \begin{equation}
           P  =  \frac{ n_S}{n} (1- (1- \oin)^{n_S})
           \label{ew_relapse}
       \end{equation} 



\subsubsection*{Experiment settings}
We run simulations with a total population of $n = 1000$ agents. These agents are connected with each other based on the pre-defined network structure. To understand how this pre-defined network structure affects smoking behaviour, we vary the network structure and study the smoking dynamics observed in each network. This process involves comparing the model on each network with the empirically observed data. Along with the comparison, we also identify the combination of parameter values that best fit observed data and how this combination changes with the underlying network. In addition, we also compare the results of the ABM on different networks with the ODE analogue (described in supplymentary section \nameref{Sup_ODE}) of the ABM.

In the ABM, each time step corresponds to a year in the real world.
At every timestep, agents follow a three-step procedure sequentially to avoid cascading agent states. First, each agent identifies a potential new state it can transition to based on the agent's state at that time step. Next, each agent calculates the probability of transitioning into the identified new state. Finally, each agent simultaneously transitions into the new state based on the calculated probability. The sequential procedure above removes the chance of cascading agent states in a single step. That is, an N-agent can never change into an S-agent and then an Q-agent in the same time step.


Due to the stochastic nature of ABMs, we iterate each simulation multiple times. We iterate each simulation ten times during the parameter sweeps for calibrating the model due to limits on computational resources. However, we iterate the best-fit combination of parameters 1000 times to validate the model, thus including possibilities of rare events.


For every simulation, a new network is generated, that is, each simulation has a different realisation of the network structure. We then initiate $s_0 \%$ and $q_0 \%$ (based on the first data-point in the empirical data) of the total $1000$ agents as S-agents and Q-agents, respectively, on the generated network.  

The model was built using the modular framework Mesa in Python \cite{python-mesa-2020}.
        
\paragraph*{Networks}

We have run simulations of our ABM on six different network topologies: fully connected, scale-free  \cite{barabasi1999emergence}, random \cite{erdHos1960evolution}, small world \cite{watts1998collective}, Lancichinetti-Fortunato-Radicchi benchmark \cite{lancichinetti2008benchmark} and a real-world network from the Framingham heart study (FHS) data \cite{hill2010infectious}. These networks were chosen due to their unique properties or ubiquitous nature in literature. The scale-free network, random network and small-world networks are standard network topologies in network sciences that are used for spreading phenomenons; hence, we test our model on these networks too. Details of each of the networks used are mentioned below.
            
                \begin{enumerate}
                    \item     Fully connected:
                    The fully connected network (FC) assumes that every agent is connected to every other agent in the system. This is to replicate the mean-field or perfect-mixing approximation seen in ODE models. However, real-world networks are sparse and seldom fully connected. Therefore, we generate and explore other network topologies.
                    \item     Random network:         
                    Similar to the FC network, every node in a random network tries to form an edge with every other node, but with a probability $p_{er}$. The situation when $p_{er}=1$ corresponds to a FC network. We use the Erdős–Rényi (ER) model \cite{erdHos1960evolution} to generate random networks for our experiments. 
                    \item     Scale-free network:
                    These are networks where the degree distribution follows a power law. Many real-world networks have been reported to follow the power-law distribution \cite{gamermann2019comprehensive, albert1999diameter}. To model this we use the Barabási–Albert (BA) network model \cite{barabasi1999emergence}.
                    
                    \item   Small-world network (SW):
                    This is a network which is highly clustered with small average shortest paths. These networks are known for local cliques and random long-ranged connections. We use the Watts–Strogatz model to generate the network \cite{watts1998collective}.
                    
                    \item     Lancichinetti Fortunato Radicchi (LFR) benchmark network: The LFR network encompasses properties of a real-world network like a heterogeneous distribution of degrees and size of communities \cite{lancichinetti2008benchmark}. We use LFR networks due to its unique property of communities embedded into it during the network generations process.
                    
                    \item Framingham Heart Study (FHS) network:
                     We have also used a real-world network based on the Framingham heart study (FHS) data along with the synthetic networks above. The FHS was a \cite{dawber2013framingham} a longitudinal cohort-based study aimed explicitly at studying cardiovascular diseases and identifying the associated factors. However, due to the wide range of documented associated factors, it has become a one-of-a-kind data set on which even detailed network analysis has been carried out. We used a configuration model to generate synthetic networks with the same degree distribution observed in the FHS \cite{hill2010infectious}. Unlike the SF and LFR networks, the configuration model allows an arbitrary distribution of degrees and is therefore not restricted to the power-law distribution \cite{newman2003structure}. We estimate two parameters for our model based on the results from the FHS data-related network analysis \cite{christakis2008collective}.
            
                    \end{enumerate}


        Each of the networks mentioned above has multiple parameters associated with its generation. We chose these parameters involved in the network generation process such that their average degree is close to the empirically observed one. From a detailed longitudinal network analysis of the FHS we know that the average degree of the network involved in the spread of smoking has an average degree of $\langle k \rangle=3$ \cite{hill2010infectious}. Based on this result, we fixed parameters involved in network generation for each of the networks mentioned above (excluding FC) to have an average degree as close as possible to 3. 
        
        The value of average-degree being close to three can be explained by the fact that the network involved in the spread of smoking behaviour mainly consists of close family members and close friends. Moreover, previous studies on the composition of social networks in other countries have returned an average degree of $4.45$ \cite{chao2015dynamic}. However, the network involved in the spread of smoking behaviour mainly consists of close family members and friends \cite{christakis2008collective}. Therefore, the number of individuals potentially influencing such behaviour will be less than the average degree of a standard social network. For this reason, $K=3$ for the generation of networks involved in the spread of smoking behaviour can be justified.
        The parameter values for each network and the average expected degree are shown in the table \ref{tab:net_par} in the supplementary material \nameref{Sup_NetPar}.

\subsection*{Data}
To calibrate and validate the model, we use publicly available smoker and quitter prevalence data from the US and UK. The UK data-set \cite{cornish2019adult} has normalised smoker population and quit ratio (defined as the proportion of smokers who have quit smoking) from 1974 to 2019. Bi-yearly data points are available until 2000, and yearly data from then on. We estimate the quitter population from the quit ratio and use it to calibrate the model.

In the case of the US, we used the data available in the official Surgeon General's report on tobacco \cite{office2020smoking}. This document reports the prevalence of smokers (male and female separately) and quitters (again, male and female separately) between the years 1965 and 2015 (data points every five years).


Additionally, we impose the values of two parameters ($\bin$ and $\gin$) of the model by estimating them from empirical research \cite{christakis2008collective}. The supplementary section \nameref{Sup_Christakis} describes the steps taken to estimate the parameters.

\subsection*{Calibration and Validation}

To calibrate and validate the model, we use a four-step process. To limit the effects of not including vital dynamics, we calibrate and validate the model on time frames of 25-30 years. Then, we split the time-stamped data into calibration and validation segments in UK and US scenarios. We use data from 1974 to 2002 (16 data points) in the calibration segment and the remaining in the validation segment in the UK data. Similarly, in the US, we use the data from 1965 - 1990 to calibrate the model and the regaining to validate it.
\paragraph*{Step 1: Coarse-grained calibration:} To identify the combinations of parameter values which best mimic the empirical prevalence data, we run a coarse-grained parameter sweep on all the uncertain parameters of the ABM. In this case, this parameter sweep was carried out for each parameter for ten logarithmically split values between 0 and 1. Further, we iterated each parameter combination ten times to reduce randomness. We then use these simulation results to identify the range of parameters best fitting the calibration data (top 100 best-fit parameter combinations). Since we have the population sizes time-series data, we use the sum of the Mean Square Error (MSE) of both the S and Q trends to identify the best fitting parameters.

\paragraph*{Step 2: Sensitivity Analysis:} Since we incorporate all possible interactions between agents, we perform a sensitivity analysis on the minimum MSE parameters to identify potentially redundant parameters. This allows us to simplify the model based on numerical simulations when possible. The detailed sensitivity analysis of each model parameter on the dynamics is available in detail in the Supplementary section \nameref{Sup_Sensitivity} 
\paragraph*{Step 3: Fine-grained calibration:}
To improve the estimated parameters, we re-calibrate the model through a finer grained parameter sweep on each parameter. In this case, this parameter sweep was carried out for each parameter for five equally split values between the range of values identified in step $1$. Just as in step $1$, we iterate each of the simulations ten times again. We use the range of values identified for each parameter from the coarse-grained calibration to run the fine-grained parameter sweep. 

\paragraph*{Step 4: Validation:}

We validate the calibrated model by comparing the simulated results with the validation data for both US and the UK. Since we used the sum of MSE of the smokers and quitters for calibrating the model, we also use the same sum of MSEs for validation. Along with the MSEs, we also use the unique crossover point of the historical trends of the smokers and quitters populations to improve the validation process.


\section*{Results}

In this section, we study the ABM for smoking and demonstrate its characteristics.
First, we compare the ABM with all possible interactions on different networks. Through this comparison, we demonstrate the importance of networks for modelling the spread of smoking behaviour. We further demonstrate the ease with which networks can be incorporated into ABMs and therefore advocate using them for modelling such a spreading phenomenon.
Next, we compare the ABM with an ODE analogue to demonstrate the equivalence of the ABM on an FC network and a traditional ODE model.
Finally, we calibrate and validate the model on empirical data observed in the US and the UK. Through the calibration and validation process, we emphasise the need to incorporate networks into such models, which can potentially be used to develop policies. Next, we show that the real-world network (FHS) replicates the empirical data observed in the US and the UK. In addition, we show that in practical situations, when complete information on the actual underlying network is not available, synthetic networks with similar average degrees can be used to develop models. On the other hand, we show that ABMs on FC networks and, therefore, ODE models should not be used for modelling smoking and similar behavioural contagion.

We examine the evolution of total prevalence of smokers (S) and quitters (Q) for each simulation setup. To compare and quantify the temporal dynamics of the populations, we calculate the sum of MSE of the S and Q curves. Since the S and Q curves cross each other in both the UK and US data, we study the unique crossover time-point for the ABM and compare it to the one from empirical data.

\subsubsection*{Population dynamics in ABMs on networks and its ODE analogue}
\begin{figure}
    \centering
    \includegraphics[scale =0.5]{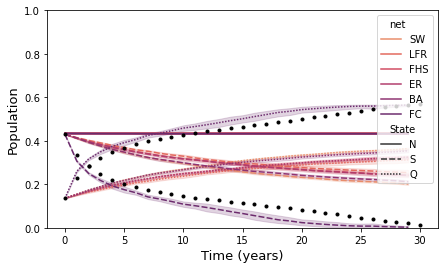}
    \caption{Dynamics of the N, S and Q populations from the ODE model and ABM on networks (FC, BA, ER, FHS, LFR and SW networks). We used the best-fit parameters from the coarse-grained parameter sweep of the FHS network  ($\oin,\din,\bspon,\gspon,\ospon = 0.01334, 0.05623, 1e-05, 4e-05, 1e-05$) for these simulations. The black lines represent the ODE model. The ODE model and ABM on FC show similar temporal dynamics, while the dynamics change drastically when the network structure moves away from FC.}
    \label{fig:FC_comparison}
\end{figure}

Figure \ref{fig:FC_comparison} shows the population dynamics observed from simulations of the ABM for smoking (on all six network topologies) and an ODE analogue (We describe the ODE model used in Supplementary section \nameref{Sup_ODE}) with the same parameter values. We run the ABM and ODE models to simulate a period of $30$ years. This time frame is similar to what we used to calibrate the models on US and UK data. We use this observed population dynamics to compare the ABM between different networks and the ABM on these networks with the ODE model.

Our results suggest that the network structure affects the population dynamics of smokers and quitters for the same experimental conditions.
Specifically, when the network structure is changed from FC to any other network, the dynamics observed change drastically. In addition, when the average degree of the networks (other than in the FC network) is kept at similar values (close to $\langle K \rangle=3$), we observe that the deviation in the observed population dynamics is minor between different networks. This deviation being minor suggests that the average degree is vital in the dynamics of smoking behaviour.

The ABM on a fully connected network follows the same trajectory qualitatively as the ODE model. Statistical equivalence of a differential equation model, which uses a mean-field approximation to an ABM on a fully connected network, has been shown before, and our model is consistent with this result \cite{rahmandad2008heterogeneity}. This suggests that a simple traditional ODE model can be used to model smoking behaviour and other similar social contagions only when the local population under study follows a fully-connected network topology. However, real-world networks are sparse and not fully connected. So ABM on FC, and by extension ODE models, should not be used to model smoking behaviour and other similar behavioural contagion processes.


\subsection*{Validation }
When the parameters that best captured the historical trend for the ABM on each network were observed, the variation in the MSE values was minor. Therefore to make the parameter selection process more robust, 
we chose 100 combinations that gave the minimum MSE values instead of choosing only the minimum MSE parameter combination. Additionally, we imposed a condition that each independent parameter in this combination had to fall between the first and third quartile of its values observed in the minimum 100 (the values which fall inside the box in Fig \ref{fig:parameter_ranges}). We then sample this new set of parameters 1000 times for validating the model.

To validate the calibrated model, we compare the evolution of simulated population sizes with the historical data. Since we calibrate the model using the MSE values of S and Q curves, we also use the same for validation. For validation, we use periods of the empirical data, which were not used to calibrate the model (validation data, periods: 2003-2019 for the UK and 1995-2015 for the USA).

\subsubsection*{Case 1: UK}

\begin{figure}
    \centering
    \includegraphics[width =\linewidth]{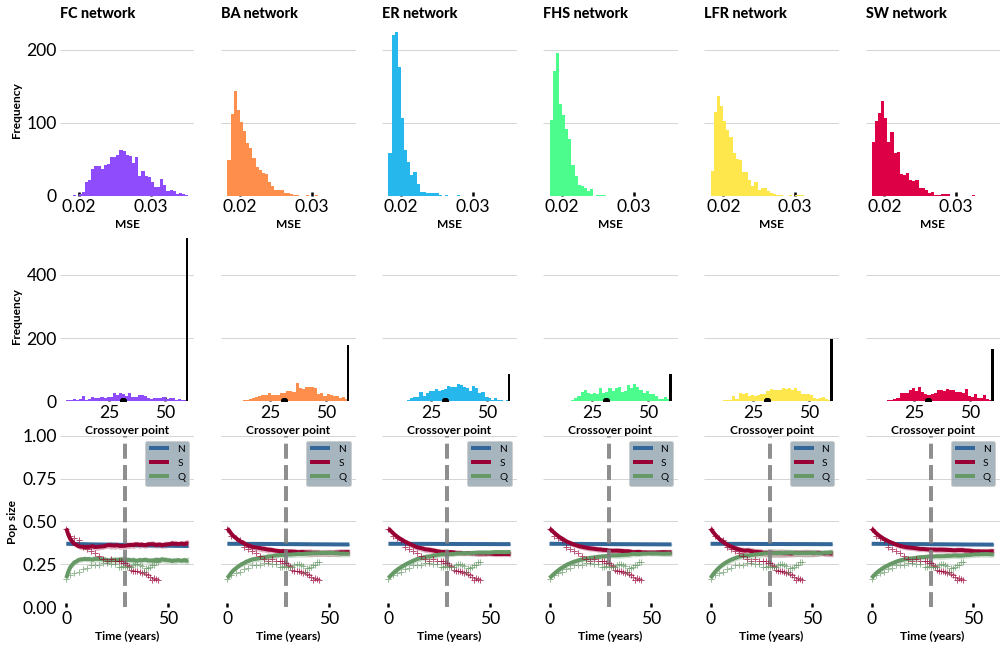}
    \caption{The figure shows simulated characteristics and population plots from 1000 runs of the ABM for the best-fit parameters on each network for the UK. The bars in the first row represent the MSE value (sum of S and Q) of ABM with the validation data, the bars in the second row show the crossover point, and the third row shows the mean population plot with a 95\% CI. The black dots on the second row on the x-axis represent the actual crossover point in the empirical data. While the black bars show the number of times the S and Q curves do not cross each other. Among the six networks, we see that ER (mean = $0.01984$, SD =  $0.00069$) and the FHS (mean = $0.01998$, SD = $0.00071$) network reproduces the data most accurately. The BA (mean = $0.02076$, SD = $0.00104$), SW (mean = $0.02081$, SD = $0.00135$) and LFR (mean = $0.02082$, SD = $0.00114$) are very similar to each other in terms how good they replicate the data. While the ABM on FC network (mean = $0.02617$, SD = $0.00226$) provides the worst fit for the validation data.}
    \label{fig:validation_UK}
\end{figure}

Figure \ref{fig:validation_UK} shows the distribution of MSE values, crossover points and the population dynamics of 1000 iterations of the simulations using the best-fit parameters for each network in the UK. The distribution of MSE projects how well the calibrated ABM on each network fits the validation data. To compare each of these MSE distribution, independent two-sample t-tests were carried out. There was no significant difference between the ABM on ER and FHS networks (p= 0.049). However, the ABM on FHS and ER were significantly different from all other networks at $5\%$ significance level. We recognise that the model has some limitations, hence we chose the alpha level of $0.05$. 
Additionally, the t-test also showed that MSE values for the LFR, BA, and SW networks were not significantly different from each other, but each of these networks was significantly different from ER and FHS networks. Importantly, There was a significant difference between FC and the other networks.

The distribution of crossover points show the value at which the crossover happens in each of the 1000 simulations (if there is a crossover). The range of crossover points observed in the distribution, always contains the actual crossover point observed in the historic trend. However, the number of times the crossover happens are very different when the underlying network is changed. The S and Q curves crosses over $485$, $822$, $915$, $912$, $802$, $836$ times for FC, BA, ER, FHS, LFR and SW respectively. The FC network both gives a higher variability in the crossover point distributions and a lower number of successful crossovers, while the ER gives the highest number of successful crossovers. From the difference in MSE values and the variability in the crossover-points we can conclude that the network structure does matter when modelling the smoking behaviour. The real-world FHS network and the ER network replicates the historic trend observed very well and also has the highest number of successful crossovers.

For the spread of smoking behaviour, the influence of other individuals is only substantial when they are a close family member or a close friend \cite{christakis2008collective}. This limits the average degree of the required network for the spread of smoking behaviour. This network will have a much lower average degree than a standard social network. In such situations, the degree distribution of the ER network can approximate real-world ones, such as that of the FHS network. This can be seen in our results, the ER and FHS network best replicate the validation data out of the six networks. The LFR and SW networks also generate low MSEs. 
This trend is seen in the number of successful crossovers as well. Following the MSE distributions, the FHS and ER also gives very high successful crossovers, while BA, LFR and SW give similar values of successful crossovers. Additionally, the ABM on FC networks gives the worst fit to the validation data and the least number of successful crossovers showing that the ABM on FC networks and, by extension, ODE models should not be used to model smoking and similar behavioural contagion. We can conclude that ABM on the ER network can be potentially used to model smoking behaviour when information on real-world network is not available.

Supplementary section \nameref{Sup_NetAnalyis} gives a detailed analysis of the networks used for the simulations.

\subsubsection*{Case 2: US}

\begin{figure}
    \centering
    \includegraphics[width =\linewidth]{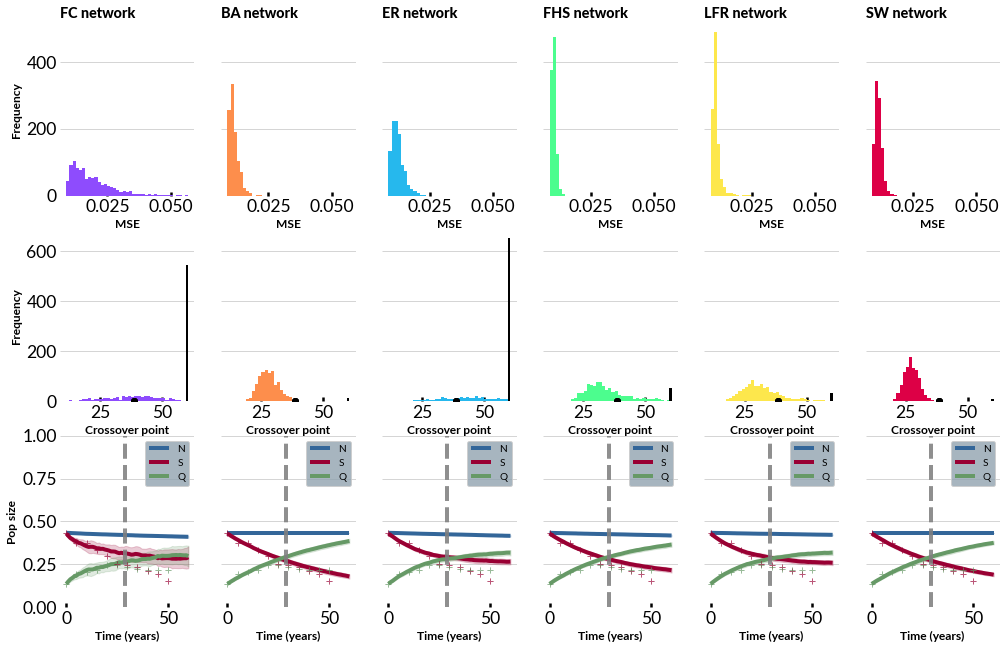}
    \caption{The figure shows simulated characteristics and population plots from 1000 runs of the ABM for the best-fit parameters on each network for the US. The bars in the first row represent the MSE value (sum of S and Q) of ABM with the validation data, the bars in the second row show the crossover point, and the third row shows the mean population plot with a 95\% CI. The black dots on the second row on the x-axis represent the actual crossover point in the empirical data. While the black bars show the number of times the S and Q curves did not cross each other. Amongst the six networks, we see that the ABM on the FHS network reproduces the data most accurately (mean = 0.01048, SD = 0.00053). The LFR (mean = 0.01102, SD = 0.00091), BA (mean = 0.0114, SD = 0.0013) and SW (mean = 0.01151, SD = 0.0007) are again very similar to each other in terms how good it replicates the data. The ER network (mean = 0.0124, SD = 0.00175) closely follows all the networks except FC. While the ABM on FC network provides the worst fit for the validation data FC (mean = 0.01878, SD = 0.00622).}
    \label{fig:validation_US}
\end{figure}

Figure \ref{fig:validation_US} shows the distribution of MSE values, crossover points and the population dynamics for the ABM over 1000 iterations using the best-fit parameters calibrated with the US data.
As in the case of the UK, we also carried out Independent two-sample t-tests for MSE distributions between each network in the US. However, there was a significant difference between the ABM on FHS and all other networks. 

Even though the distribution of MSE values were significantly different, the average MSE values of the ABM on each of the networks were lower than that of the ABM on FC and very close to each other. We find that the FHS network fits the real-world data the best in the case of the US, followed by the LFR network. Opposite to the others, the LFR network has a unique property of communities structure embedded in the network generation process. This suggests that communities play a role in the spread of smoking behaviour.

\subsubsection*{Analysis of Best-fit parameters}
\begin{figure}
     \centering
     \begin{subfigure}[b]{\linewidth}
         \centering
         \includegraphics[width=\linewidth]{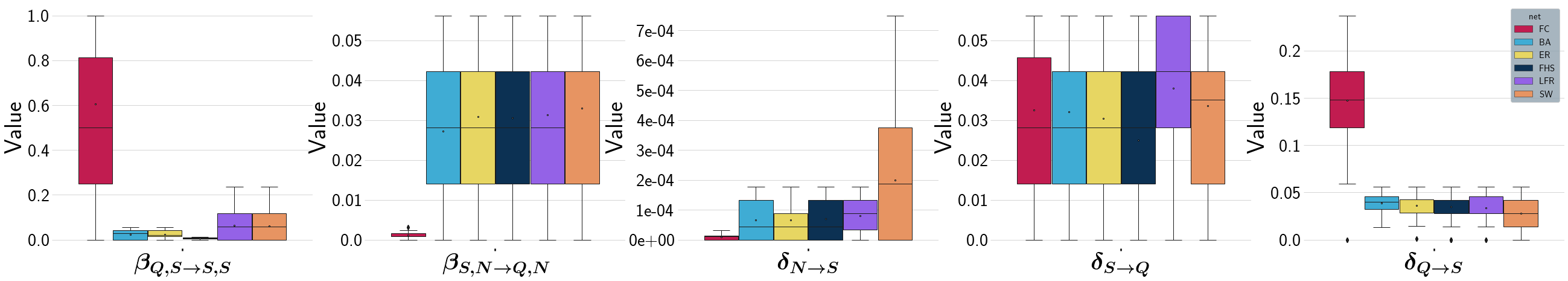}
         \caption{UK data}
         \label{fig:y equals x}
     \end{subfigure}
     
     \hfill
     
     \begin{subfigure}[b]{\linewidth}
         \centering
         \includegraphics[width=\linewidth]{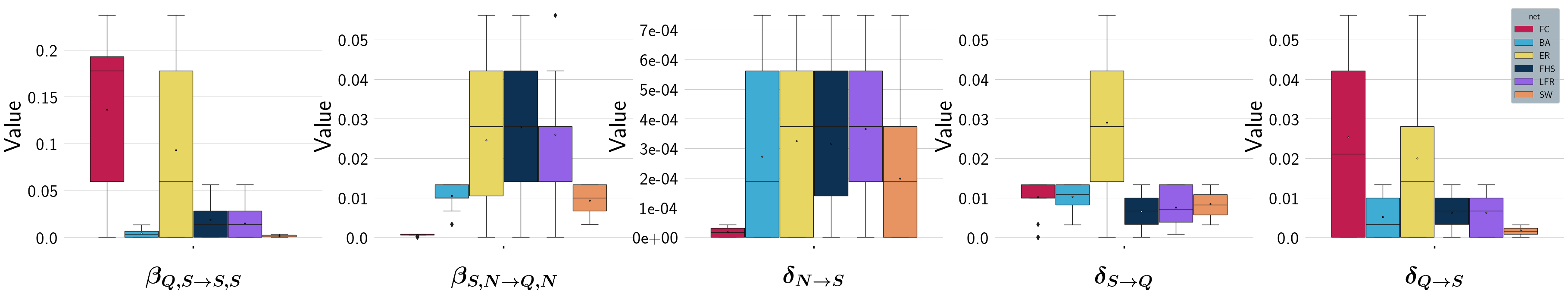}
         \caption{US data}
         \label{fig:five over x}
     \end{subfigure}
        \caption{Box plot representing the range of values for each parameter in the 100 simulations, which best fit the calibration data for each network. The white circle in each box shows the mean value of the parameters. From left to right, we have the FC (red), BA (light blue), ER (yellow), FHS (dark blue), LFR (purple) and SW (orange) networks for each plot. }
        \label{fig:parameter_ranges}
\end{figure}

Figure \ref{fig:parameter_ranges} shows a box plot of values seen in the 100 best-fit parameters for each network on both US and UK data.
When the 100 best parameter combinations that fit the data best were compared, the FC network consistently gave significantly different parameter combinations in both US and UK data sets. In both US and UK, at the $5\%$ significance level, the values of at least 80\%  of the parameters returned by the FC network were significantly different from the other networks. Moreover, 100\% of them were significantly different from the FHS network.
On the other hand, all other networks return parameters in which at least 20\% of them are not significantly different from that of the FHS. The only exception is the SW network in the case of the US, where all parameters returned were significantly different.

The ER and BA networks perform well in the case of the UK (almost 60\% of the parameters are not significantly different from the FHS network), but in the case of the US, similarity of the parameters drops (only 40\% are not significantly different). However, the LFR network performs decently in the UK data set (60\% are not significantly different) and very well in the US case (none of the parameters is significantly different). 

Our results thus indicate that when the average degree is kept constant, the parameters returned by the FHS network are somewhat similar to that of the LFR, ER and BA networks. 

By comparing the ABM calibrated on different networks between the US and the UK, we see that in the LFR, SW, and FHS networks,  20\% of the parameters were not significantly different between the US and UK. While in the ER network, 40\% of the parameters were not significantly different. However, moving from the UK to the US in the BA and FC network, all the parameters returned after the calibration process significantly differed. This shows that the ABM on FHS, ER, SW and LFR networks are robust despite geographic variations. However, this is not true for the FC and BA networks.

\paragraph*{Implications to policy}
When the underlying network structure is changed, the parameters best replicating the empirical data also change. Therefore, when models are used to develop policies, parameter estimation becomes very important to predict the outcome of potential new policies. If the wrong network structure is used for the model, the calibrated parameters will also be different, which would lead to inaccurate strategies being developed. 
However, our results (robustness of the FHS, ER, SW and LFR  to changes in geographic regions and the similarity of parameters of LFR, ER and BA within each region) suggest that when the real-world network structure is not available (in our case, the FHS network), the LFR network or the ER network provides a satisfactory approximation. 
Thus, LFR or ER networks could potentially be used to develop strategies for controlling smoking behaviour when the local population's underlying network structure is unavailable.

\section*{Discussions}
We have developed a network-based ABM to study the spread of smoking behaviour. We need such a model to develop policies, so all possible interactions between individuals are considered. Recent empirical studies have pointed out how interactions between individuals influence smoking behaviour \cite{christakis2008collective,blok2017role,ennett2008peer,mercken2009dynamics}. However, this has not been incorporated completely into the ODE models developed to keep the model less complex (and solvable). However, using ABMs, such interactions can be easily modelled without raising the complexity of the model.

When we incorporate interactions into a model for smoking, we quickly realise that not every individual in a society influences everyone else. Only close family members and friends can influence smoking behaviour substantially \cite{christakis2008collective}. Therefore it is necessary to consider the underlying network of close friends and family for such modelling smoking. Not considering network structure would imply using an assumption similar to the well-mixed approximation in ODE models. However, our results suggest that ODE models perform very poorly compared to ABMS on networks.

Our ABM on FC network performs very poorly compared to other networks. In the ABM for FC network, all agents are connected to each other. This makes it equivalent to the ODE model's well-mixed or mean-field approximation. Only in situations where the entire population is closely connected, the ODE model can be used without considering the network structure in place of the ABM. However, this is seldom the case in real-world networks, and therefore ODE models which assume a well-mixed population and ABMs on FC networks should not be used for modelling smoking behaviour.

 By considering other network topologies, we move away from this mean-field approximation, thus resulting in different dynamics from the ODE model. In addition, real-world networks are also sparse networks; because of this, dynamics observed on other networks will not be similar to that of the ODE model, as seen from our results.

Using our ABM, we study the effect of network topology on the dynamics of the spread of smoking and see that the network structure affects it. The effect of network topology can be clearly seen when moving from an FC network to other networks. However, this difference is minor within other networks when the average degree is at similar values. This suggests that, in addition to the network structure, the average degree of the underlying network might also play a role in the spread of smoking behaviour. 

Our UK and US results suggest that ER and LFR networks replicate the empirical data better than the other networks (excluding FHS). Apart from this, from our analysis of parameters returned during calibration, we observe that only ER and LFR networks are robust to changes in the geographic region and also return a combination of parameters of which at least $40\%$ are not significantly different from the FHS network. 
On closer observation of the network characteristics of LFR, ER and FHS networks (Figure \ref{fig:net_analysis}), we see that all three of the networks have very similar average degrees and thus also form a similar number of edges. At the same time, the BA and SW have the same average degree as each other, which is a bit higher than the other networks. This suggests that the average degree plays a crucial role in the dynamics of smoking behaviour and possibly, any synthetic network with an average degree similar to that of the real-world network can be used to model smoking behaviour.

We have demonstrated through the calibration and validation process that the ABM on networks can be used to model smoking behaviour.
Even with the wrong network,  these models may estimate the trajectory of the populations qualitatively but only with a lower accuracy. Nevertheless, when a network changes, the parameters returned on calibration also change. 
The differences in the model and its best-fit parameters become important when policymakers use the predictions to develop strategies to curb smoking. Incorrect approximation of network structure and thus the model parameters can potentially lead to the development of ineffective policies.

However, our results suggest that in cases where the real-world network information is not entirely available due to practical constraints, policymakers can use LFR networks and ER networks, as they replicate the empirical data with good accuracy and return parameters which are not significantly different from the real-world network. The ineffective strategies that policymakers would have made without considering networks or the wrong network can be improved in efficiency by using ER or LFR networks. The development of strategies to control smoking based on the network structure is beyond the scope of this paper, and we leave it as a future task.

    \paragraph*{Limitations}
    
 ABM is a good technique for incorporating network structure and interactions between individuals to study the macroscopic outcome. Our model shows that the network structure of the population is essential while modelling smoking, which should be taken into consideration while developing policies. However, some limitations due to modelling choices to preserve the model's simplicity should be noted.

 First, we have assumed that every individual behaves in the same way within a group. However, this is not the case in a real-world setting.

Second, we have not considered vital dynamics in the model and the age-dependent nature of reaction to influence. Usually, a constant mortality rate and birth rate are incorporated to model vital dynamics. However, in network models, there is a risk of older agents (initial agents when the model is initiated) gaining more centrality just because of the implementation of network growth. A careful understanding of age-dependent mortality rates and social-ties should be incorporated into the model to circumvent this problem. However, this is beyond the scope of this paper and will be explored in future work. Therefore, to limit the effects of not including vital dynamics into the model, we calibrate and validate the model in time-periods of 25-30 years.

 Third, the degree of influence on smoking behaviour changes with the social tie you have with the other person \cite{christakis2008collective}. As a starting point in modelling influence on smoking behaviour, we have assumed that every kind of relationship affects the smoking behaviour similarly. Further, many social ties can be one-directional (as perceived by one individual in a social tie). However, we have not considered any directed graphs in the model.

Since our model assumes that the total population is constant, $N = 1- Q - S$, no new N-agents are being introduced to the system. Additionally, the S curve is calibrated against a decreasing trend observed in the empirical data, which explains the absence of variation NS population.

\section*{Conclusions}
We have developed an agent-based model for smoking dynamics that considers the contagious nature of smoking behaviour by including network effects. This model can act as a test-bed for network based policies and strategies to control the spread of smoking behaviour.

Our results suggest that, when interactions between individuals are used to model population-level smoking dynamics, the underlying network of the local population becomes very important. By changing the network topology from a fully connected network to other theoretical networks and, finally, a real-world network, we show that the dynamics deviate drastically from those of a traditional ODE model. 

We show that our model is robust and consistent to the historic trends observed in two countries - US and UK. In both countries, the network based on a real-world local population best replicates the trends observed.

Moreover, our results suggest that, the network topology, the average number of social ties and the presence of communities in the population improve the accuracy of the model.

Importantly, given the difficulties in collecting data on offline social networks, we find that the LFR and ER network replicates the empirical data with considerable accuracy and also returns parameters which are not significantly different from those of the FHS network, suggesting that the LFR and ER networks can be used for models of tobacco use.
\section*{Acknowledgments}

This work has made use of the resources provided by the Edinburgh Compute and Data Facility (ECDF) (http://www.ecdf.ed.ac.uk/).

\bibliography{biblio.bib}

\newpage
\section*{Supporting information}

\paragraph*{S1 Appendix}\label{Sup_binomial}
\textbf{Binomial approximation for independent exposure}

 Suppose agent $i$ with state Z has $n_i$ neighbours, of which  $n^{Y}_i$ are in state Y. Let X be the event in which the agent $i$ interacts with a Y neighbour, this can have two outcomes, success = (agent $i$ becomes a Y agent), or failure = (agent $i$ stays a X agent). Let $b$ be the probability of successful interaction of agent $i$ with the Y agent and k be the number of successes, then the probability of k successes is:

 \begin{equation}
    P(k ) = {\binom{n^{Y}_i }{k}} b^k (1-b)^{(n^{Y}_i-k)}
\end{equation}

Probability of state change is then:
\begin{align}
    P &=  P ( k> 0 ) \\
    &= 1 - P ( k =0 )  \\
    &= 1- (1-b)^{n^{Y}_i} 
\end{align}

This value  only focuses on the total number of Y agents and doesn't take into account the total number of neighbours the agent $i$ has.  \cite{blok2017role} shows that more than the number of individuals in a particular state (smokers or quitters), it is the proportion of them in your relationship circle that has an impact on an individuals smoking behaviour. In order to incorporate the size of the network neighbours, we scale it with the fraction of Y neighbours to the total number of neighbours. Therefore the probability for smoking initiation due to interaction is

\begin{equation}
    \bin= \frac{n^{S}_i}{n_i} (1 - (1-b)^{n^{S}_i})
    \label{bin_init}
\end{equation}

Similarly, the probability of smoker agent $i$ quitting due to interaction with quitters is:

\begin{equation}
\gin= \frac{n^{Q}_i}{n_i} (1 - (1-g)^{n^{Q}_i})
    \label{bin_ces_Q}
\end{equation}

Probability of smoker agent $i$ quitting due to interaction with never-smokers is:

\begin{equation}
\din= \frac{n^{N}_i}{n_i} (1 - (1-d)^{n^{N}_i})
    \label{bin_ces_S}
\end{equation}

Probability of quitter agent $i$ relapsing due to interaction with smokers is:
\begin{equation}
\oin= \frac{n^{S}_i}{n_i} (1 - (1-o)^{n^{S}_i})
    \label{bin_ces_I}
\end{equation}

\paragraph*{S2 Appendix}\label{Sup_ODE}
\textbf{ODE model}
            
    Compartmental models are differential equation models in which the entire population is divided in to different compartments. In our case, each individual in the population is part of one of the three compartments: Never-smoker, Smoker or Quitter.
    Similar to the popular SIR model, our model also divides the population into three compartments. However, unlike the traditional SIR model, in our case, both infection (smoking initiation in our case) and recovery (quitting in our case) are contagious. In addition we consider relapse and never-smokers influencing smokers too. These are governed by the following equations.

    \begin{equation}
      \dot{N}=  -\frac{\bin }{n  } N  S -   \bspon S
    \end{equation}

    \begin{equation}
        \begin{aligned}
            \dot{S}&=  \frac{\bin}{n  } N S -\frac{\din}{n  } S  N-\frac{\gin }{n  }S    Q  \\ 
            & +\frac{\oin}{n  } Q    S  + N\bspon +  Q\ospon - S \gspon 
        \end{aligned}
    \end{equation}
    
    \begin{equation}
        \begin{split}
            \dot{Q} & = \frac{\din }{n  }S  N + \frac{\gin}{n  } S    Q - \frac{\oin }{n  }Q    S \\ 
            & +   N\bspon +Q\ospon -S \gspon 
        \end{split}
    \end{equation}

    The equations above models the temporal evolution of each of the compartments: Never-smokers $N(t)$, smokers $S(t)$, and quitters $Q(t)$. At any point in time, the total population $n = N + S +Q$. 
    The interaction parameters are given by $\beta$'s with the subscripts showing the interacting states and the spontaneous terms are given by $\delta$'s with the subscripts showing the direction of flow of individuals between compartments.
    
    As mentioned in the main section of the paper, the ODE model assumes a perfect mixing situation where any individual can interact with any other individual in the population. 
\paragraph*{S3 Appendix} \label{Sup_Christakis}
\textbf{Parameter estimation from Christakis}

The paper by Christakis et al. \cite{christakis2008collective} quantifies the relative probability of an individual picking up smoking and quitting smoking due to interaction with a close social tie using the Framingham heart study (FHS) data set. We use the values from the paper to calculate the interaction parameters $\bin$ and $\gin$.

The FHS study is a longitudinal observational cohort study initiated to study various risk factors involved with cardiovascular diseases. The Christakis paper looks at 7 exam waves and calculates the relative probabilities using the method of logistic regression. The paper gives the relative increase in probability of smoking initiation and cessation due to interaction. According to our definitions in the model, this implies that the interaction probabilities for smoking initiation and smoking cessation are related to their respective spontaneous terms.

Since our model does not differentiate between social relationships, we take a weighted average of the relative probability considering the total number of members falling in each of the different social ties mentioned in the paper. We consider the average probability from the seventh wave (the final one) for our calculations.\\
 From the paper, we have calculated these values as
  \begin{equation}
        \begin{aligned}
            g & = 0.35214  \\ 
            b & = 0.40719
            \end{aligned}
    \end{equation}
    Where $g$ and $b$ are the weighted average of the probabilities of smoking cessation and initiation respectively (averaged considering different social tie).\\
As mentioned in the main paper, our model used a time unit of 1 year. However, the Christakis paper calculates the probabilities based on a regression between seven unequally split examinations (unequal in the time space). 
The detailed derivation of the steps used to calculate the yearly probabilities is given below.

\textbf{Calculating yearly probabilities from probability over multiple years}\\
Suppose the initial population of smokers is $S_0$ and the probability of quitting in a year is $p$. Let,
$Q_t$ be the number of people who will quit in year $t$ and $S_t$ be the number of people remaining in smoker compartment in year t.

Then,
\begin{equation}
    \begin{aligned}
        S_1 &= (1-p) S_0 \\ 
        S_2 &= (1-p)^2 S_0  \\
        &\vdots\\
        S_t &= (1-p)^t S_0
    \end{aligned}
\end{equation}
Therefore the number of quitters at every time step becomes,
\begin{equation}
    \begin{aligned}
        Q_1 &= p S_0   \\
        Q_2 &= p S_1 = p(1-p)S_0  \\
        & \vdots\\
        Q_t &= p S_{t-1} = p(1-p)^{t-1} S_0
    \end{aligned}
\end{equation}

The total number of quitters due to this state change becomes
\begin{equation}
    \begin{aligned}
        Q_{total}&= \sum_{i=1}^{t} Q_i \\
        &= p S_o(1+  p(1-p)  + p (1-p)^2 + \hdots + (1-p)^{t-1})\\
        &= p S_0  \sum_{i=1}^{t} (1-p)^{t-1}\\
        &=S_0 (1- (1-p)^t)
    \end{aligned}
\end{equation}

The paper reports the probability of quitting over $t$ years which is equivalent to:
\begin{equation}
    \begin{aligned}
       g &= \frac{Q_{total}}{S_o} \\
       &= 1-(1-p)^t
    \end{aligned}
\end{equation}

Therefore the probability of quitting per year is,
\begin{equation}
    p= 1- (1- g)^{1/t}
    \label{t_to_1year}
\end{equation}

This can be generalised for both smoking initiation and smoking cessation. 

By applying \ref{t_to_1year} to both the interaction term and spontaneous term, we can calculate the interaction probability per year as
\begin{equation}
    \begin{aligned}
        \gin & = 1 - \left[ 1- g (1- (1- \gspon)^t     \right]^{1/t} \\
        \bin & = 1 - \left[ 1- b (1- (1- \bspon)^t     \right]^{1/t}
    \end{aligned}
\end{equation}

We assume that, each person got examined in the same order, and take the average of the time between the starting year of each examination to calculate the time between two exams. We estimate the value of $t=4.5$
\paragraph*{S4 Appendix}\label{Sup_NetPar}
\textbf{Parameters used for network generation }

Table \ref{tab:net_par} shows the parameter values used to generate networks in each run of the ABM. Each of the parameters were chosen, so that the average degree of the network, was as close as possible to the average degree observed in the FHS network. For LFR networks, we used the standard parameters used for bench-marking \cite{lancichinetti2008benchmark} along with added condition of the average degree.

\begin{table}
\centering
\begin{tabular}{|c|c|c|}
\hline
Network & Parameters    & Average degree       \\ \hline
BA        & $m_{BA} = 2$      &  $4$           \\ \hline
ER        & $p_{ER} =0.003$      &  $3.1$           \\ \hline
SW        &      $k = 4$, $p = 0.3$             &   $4$          \\ \hline
FHS       &         degree distribution          &     $3.0$        \\ \hline
LFR       &     $\tau_1=2.5$, $\tau_2=1.5$,  $\mu =0.6$, $<k>=3$             &        $3.0$     \\ \hline
\end{tabular}

\caption{Model parameters involved in network generation}
\label{tab:net_par}
\end{table}
\paragraph*{S5 Appendix}\label{Sup_ParO}
\textbf{Effect of parameter $\oin$ in ABM on FC in the UK }

\begin{figure}[h]
    \centering
    \includegraphics[scale=0.3]{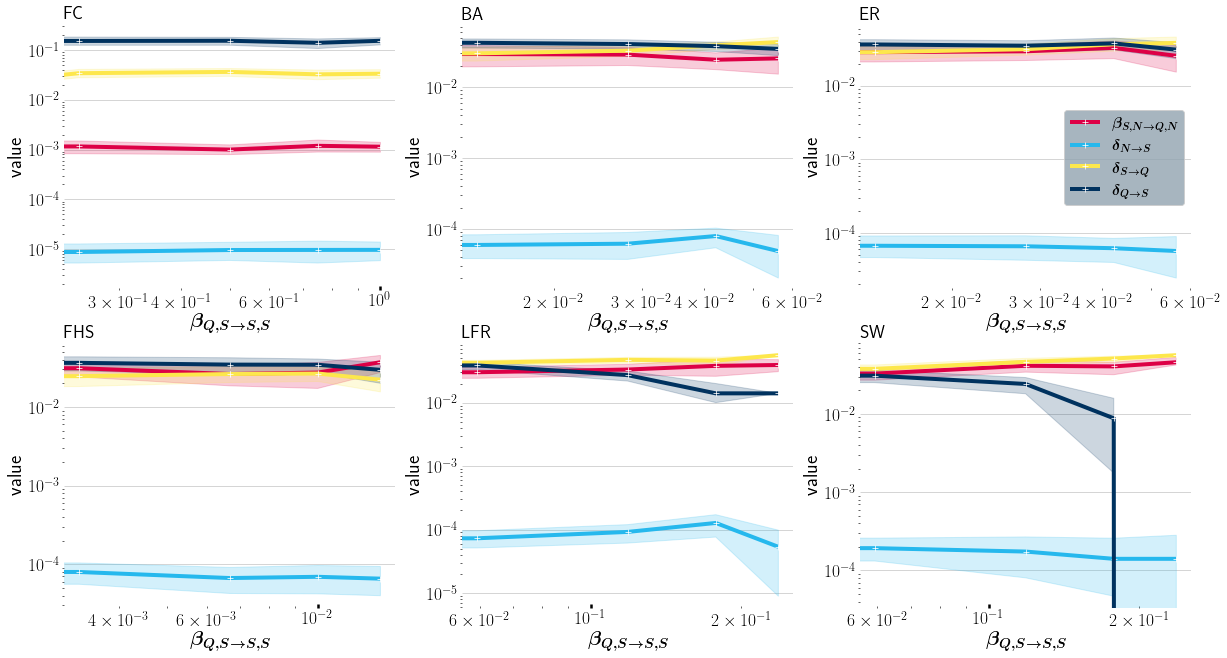}
    \caption{Effect of parameter $\oin$ on other parameters in the 100 best-fit parameter combinations. In clockwise direction, we have FC, BA, ER, SW, LFR and FHS networks.}
    \label{fig:parameter_o_vs_other}
\end{figure}

Figure \ref{fig:parameter_o_vs_other} shows how the their parameters change in each of  the network, when the parameter $\oin$ changes. We see that the parameter $\oin$ (interaction based relapse) doesn't have any effect on the other parameters in the 100 best fit parameters. This along with the fact that the range of parameters (range = 0 to 1) we see in the best 100 parameters shows us that changing the value of the $\oin$ doesn't affect the accuracy of the model in mimicking the empirical data for the ABM on FC network in the UK. However, this is not observed in the case of FC network in the US.

In the case of other networks, $\oin$ does not make much of a difference except in the case of SW network ($\ospon$). However the value of $\oin$ doesn't span the entire range of values available, so we can't conclusively say that parameter $\oin$ is redundant in the other cases. 
\paragraph*{S6 Appendix}\label{Sup_NetAnalyis}

\textbf{Network analysis}

\begin{figure}
    \centering
    \includegraphics[scale=0.3]{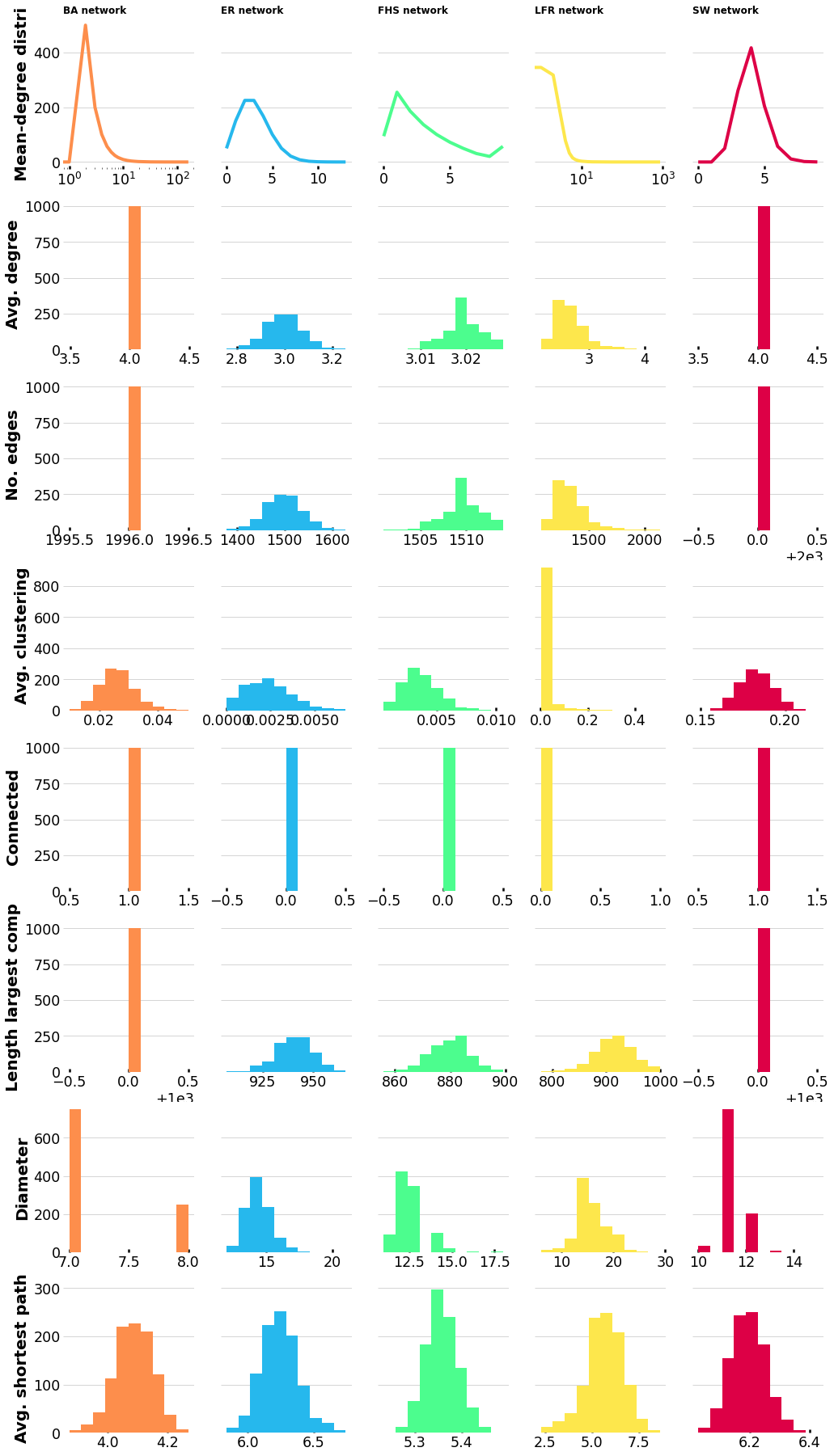}
    \caption{The figure shows the average network characteristics of the 1000 simulated networks used for validation for each network topology. Each column represents a different network topology, while each row gives the corresponding network characteristic.}
    \label{fig:net_analysis}
\end{figure}

Figure \ref{fig:net_analysis} shows the average network characteristics of the 1000 networks used for running the ABM on all network topologies other than the FC network. The first row depicts the degree distribution averaged over 1000 iterations. Not surprisingly, only the BA and LFR networks produced networks with a high degree heterogeneity, while the FHS, ER and SW networks have a similar range of degrees in the distribution. The second row shows the distribution of average degree, which by design is around the value $\langle k \rangle=3$ in ER, FHS and LFR. While, in the case of BA and SW it is around the value of $4.0$.
Only the BA and SW networks produced completely connected networks. The other three networks produced disconnected networks with varying sizes of the giant component.
The BA network has the shortest diameter and the shortest average path, while ER and FHS networks have the smallest average clustering coefficient. Comparing the multiple network statistics of the generated networks, we can see that the ER network is very closely similar to the real-world FHS network.

\begin{figure}
    \centering
    \includegraphics[width=\textwidth]{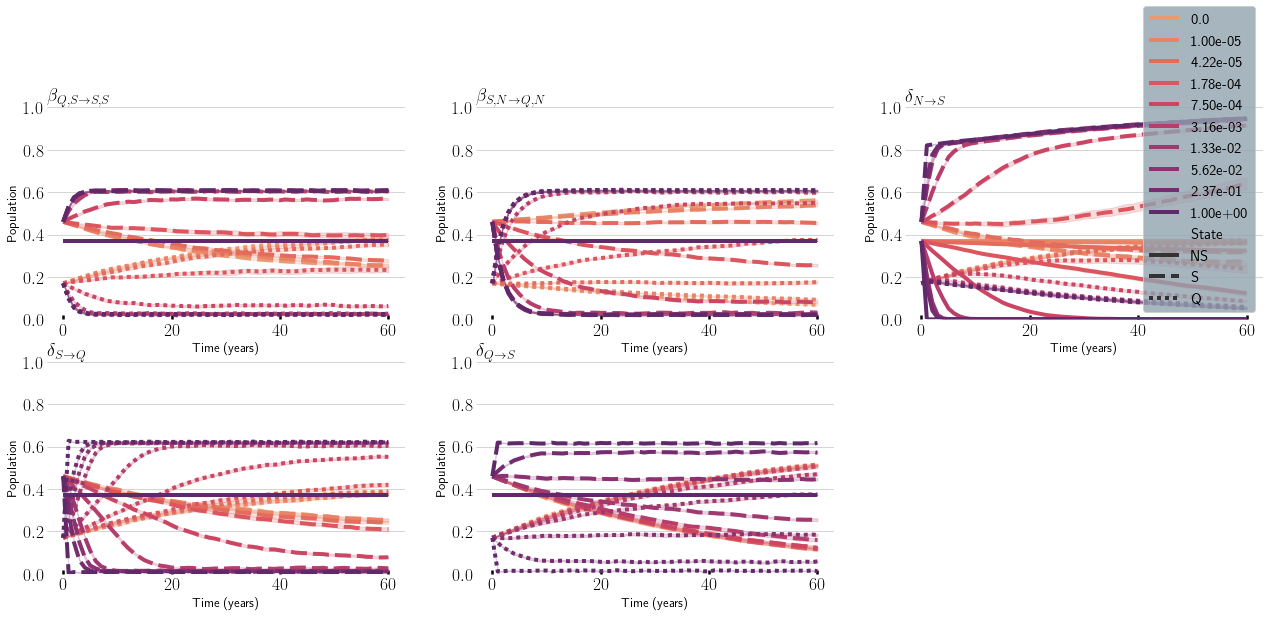}
    \caption{Sensitivity analysis of parameters around the best-fit parameters for the ABM on FC. From left to right, row by row, the effect of increasing parameters $\oin$, $\bin$, $\bspon$, $\gspon$, and $\ospon$ on the population dynamics is displayed. Each parameter is increased from $0$ to $1$ (10 values) logarithmically, while all other parameters are kept at the minimum MSE parameter combination.}
    \label{fig:SA_FC}
\end{figure}

\paragraph*{S7 Appendix}\label{Sup_Sensitivity}
\textbf{Sensitivity Analysis}
From Figure \ref{fig:SA_FC} we can see that by increasing the parameters $\oin$, $\bspon$, and $\ospon$, the value S population while the Q population decreases. These are the parameters that directly increase S population size, and thus the result can be logically reasoned from the model structure. At the same time, by increasing the parameters $\gspon$ and $\din$, the Q curve increases, and the S curve decrease. 
Only $\bspon$ affects the N curve, which is the parameter directly involved with smoking initiation. Since the N population is not affected much, the S and Q curves complement each other in all other cases.

\begin{figure}
    \centering
    \includegraphics[width=\textwidth]{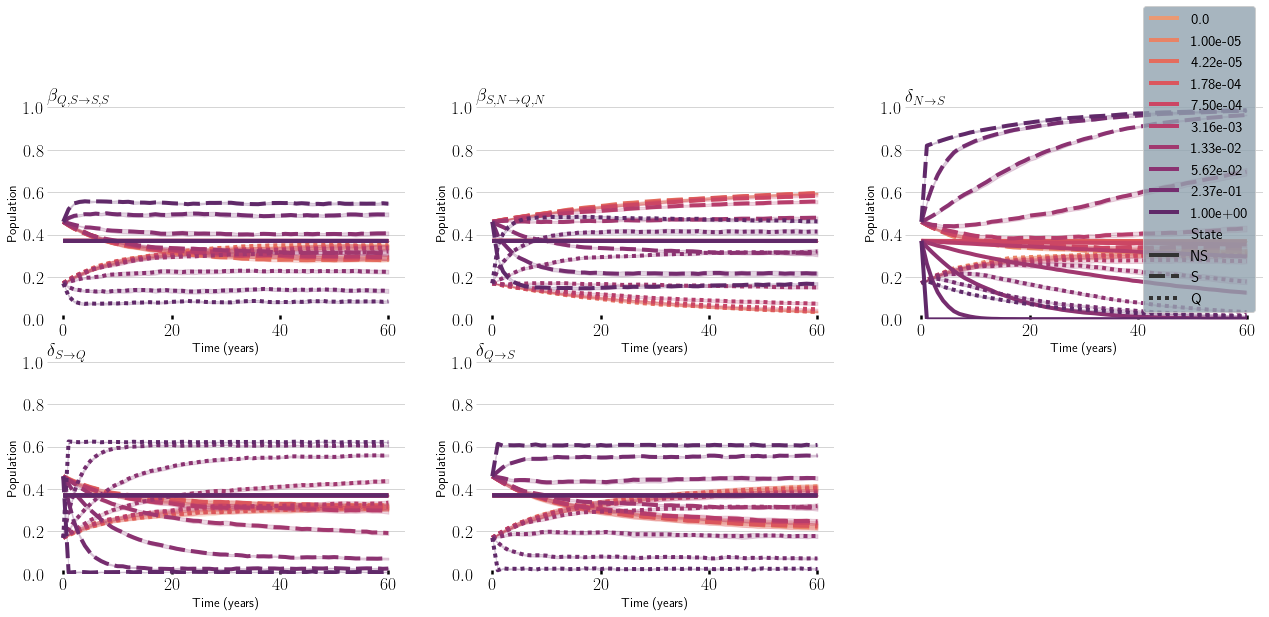}
    \caption{Sensitivity analysis of parameters around the best fit parameters for the ABM on BA. From left to right, row by row, the effect of increasing parameters $\oin$, $\bin$, $\bspon$, $\gspon$, and $\ospon$ on the population dynamics is displayed. Each parameter is increased from $0$ to $1$ (10 values) logarithmically, while all other parameters are kept at the minimum MSE parameter combination.}
    \label{fig:SA_BA}
\end{figure}

\begin{figure}
    \centering
    \includegraphics[width=\textwidth]{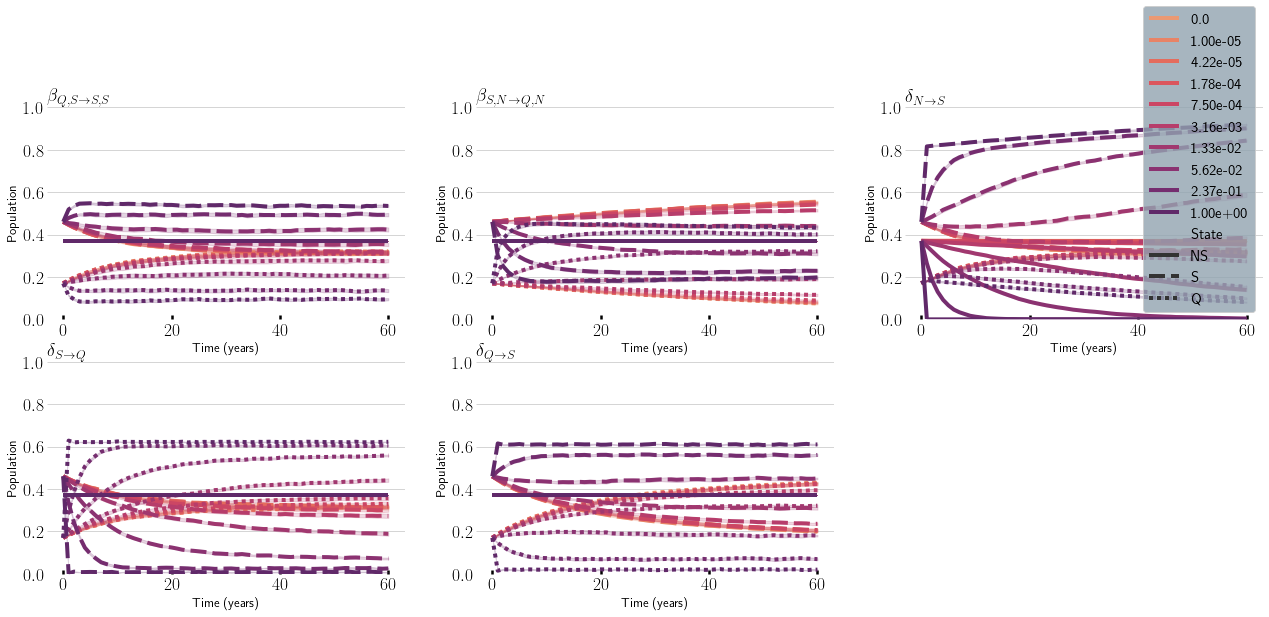}
    \caption{Sensitivity analysis of parameters around the best fit parameters for the ABM on ER. From left to right, row by row, the effect of increasing parameters $\oin$, $\bin$, $\bspon$, $\gspon$, and $\ospon$ on the population dynamics is displayed. Each parameter is increased from $0$ to $1$ (10 values) logarithmically, while all other parameters are kept at the minimum MSE parameter combination.}
    \label{fig:SA_ER}
\end{figure}

\begin{figure}
    \centering
    \includegraphics[width=\textwidth]{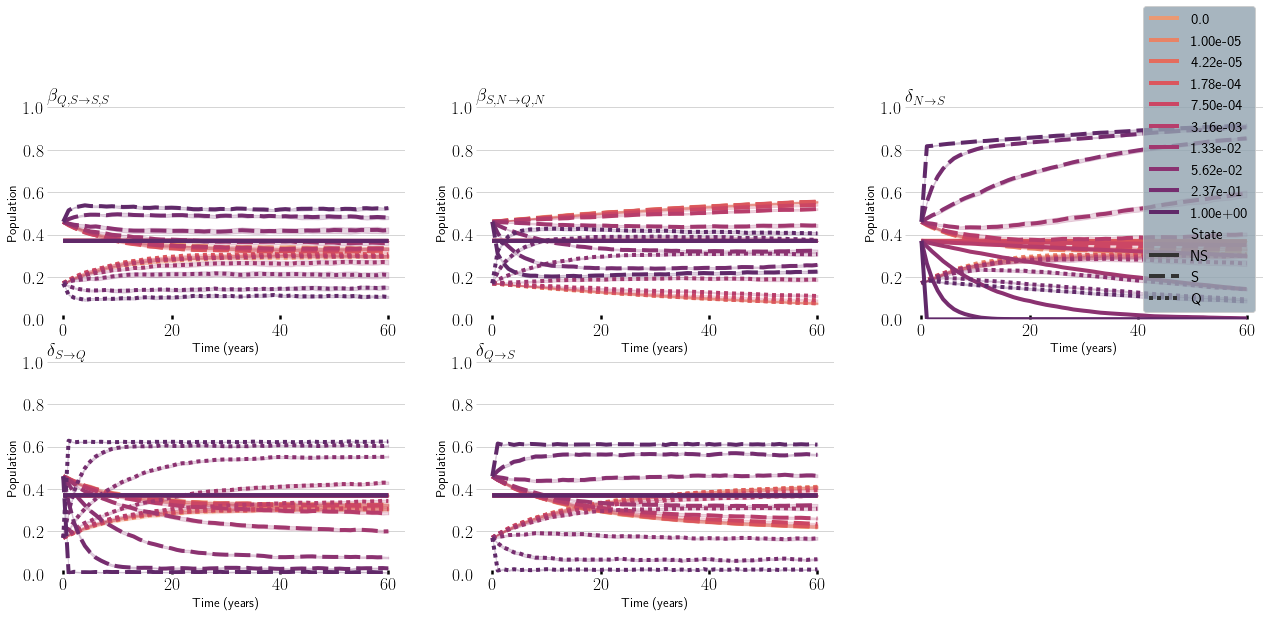}
    \caption{Sensitivity analysis of parameters around the best fit parameters for the ABM on FHS. From left to right, row by row, the effect of increasing parameters $\oin$, $\bin$, $\bspon$, $\gspon$, and $\ospon$ on the population dynamics is displayed. Each parameter is increased from $0$ to $1$ (10 values) logarithmically, while all other parameters are kept at the minimum MSE parameter combination.}
    \label{fig:SA_FHS}
\end{figure}

\begin{figure}
    \centering
    \includegraphics[width=\textwidth]{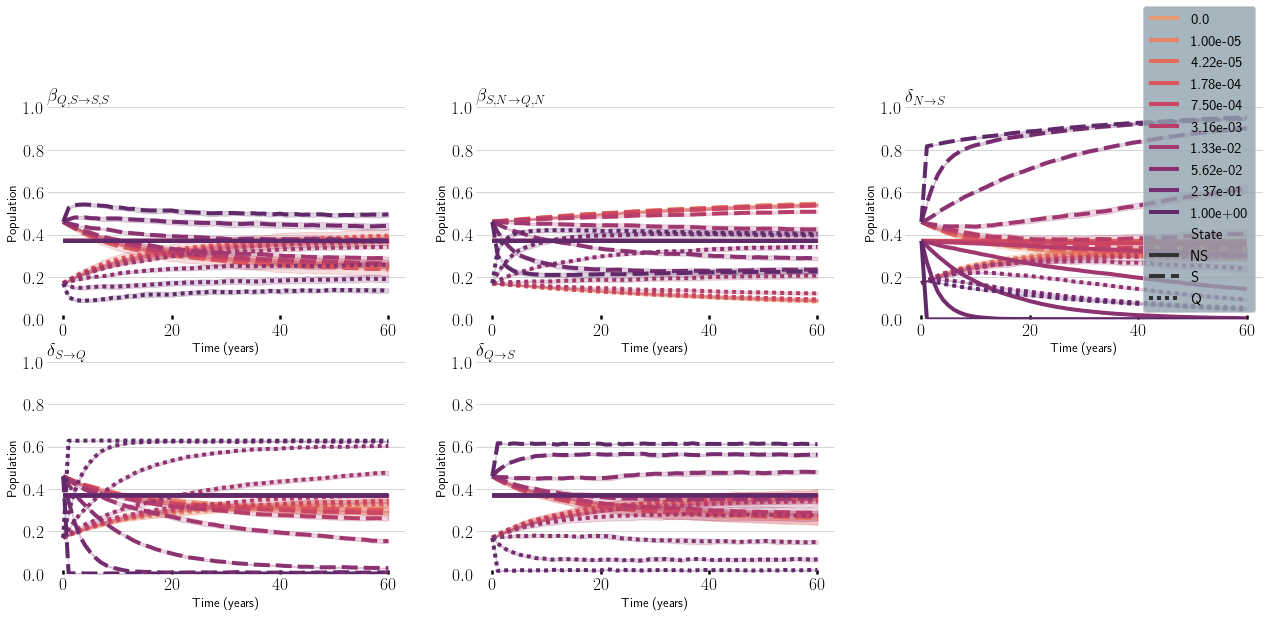}
    \caption{Sensitivity analysis of parameters around the best fit parameters for the ABM on LFR. From left to right, row by row, the effect of increasing parameters $\oin$, $\bin$, $\bspon$, $\gspon$, and $\ospon$ on the population dynamics is displayed. Each parameter is increased from $0$ to $1$ (10 values) logarithmically, while all other parameters are kept at the minimum MSE parameter combination.}
    \label{fig:SA_LFR}
\end{figure}

\begin{figure}
    \centering
    \includegraphics[width=\textwidth]{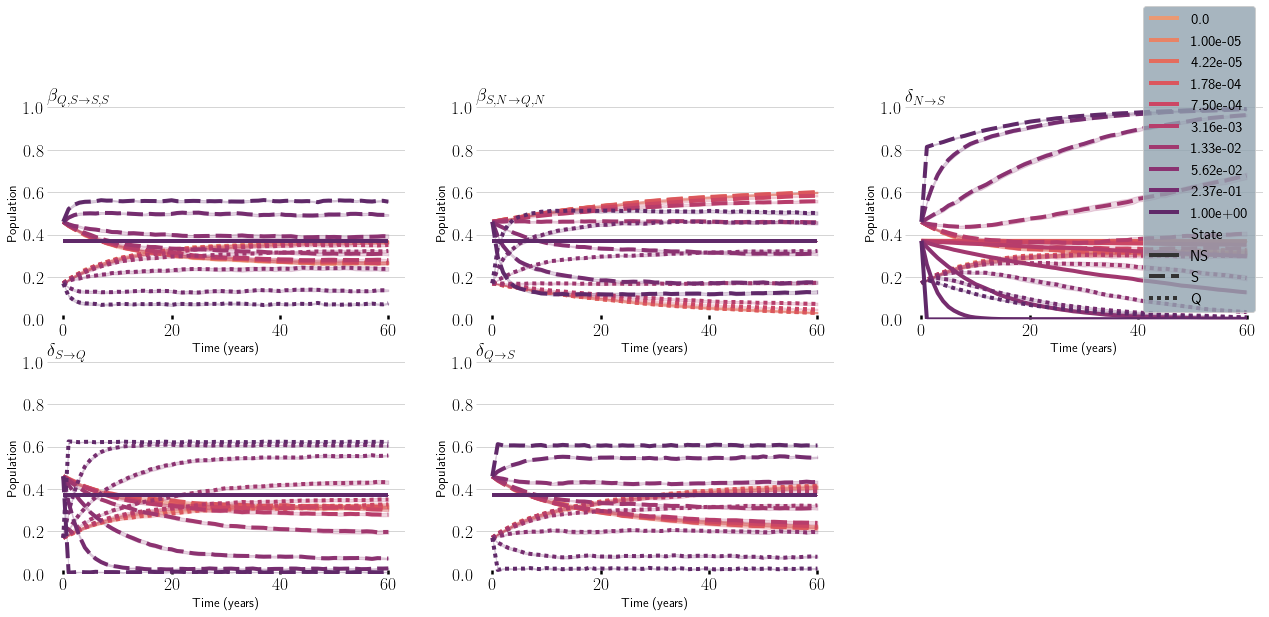}
    \caption{Sensitivity analysis of parameters around the best fit parameters for the ABM on SW. From left to right, row by row, the effect of increasing parameters $\oin$, $\bin$, $\bspon$, $\gspon$, and $\ospon$ on the population dynamics is displayed. Each parameter is increased from $0$ to $1$ (10 values) logarithmically, while all other parameters are kept at the minimum MSE parameter combination.}
    \label{fig:SA_SW}
\end{figure}

Just as in Figure \ref{fig:SA_FC}, the other networks also behave similarly in terms of how each parameter affects the population dynamics. This can be seen in Figures \ref{fig:SA_BA} and \ref{fig:SA_ER}, \ref{fig:SA_FHS}, \ref{fig:SA_LFR}, \ref{fig:SA_SW}. However, the degree to which each parameter affects the dynamics differs, especially between FC and the other networks. Our results from the Sensitivity analysis show that the FC network is more sensitive to lower values of parameters for all parameters except $\ospon$. At higher values of most of these parameters, the population behaviour saturates and converges to a similar value. 

In the case of all the other networks, the behaviour displayed on changing parameters around the best-fit combination is very similar. This is probably due to each of these networks having similar average degrees.

\newpage

\end{document}